\documentclass[prl,twocolumn,aps,amsfonts,superscriptaddress,showpacs]{revtex4}
\usepackage{bm}
\usepackage{graphicx}

\def\bsigma{{\bm{\sigma}}}
\def\b{{\bf{b}}}
\def\db{\delta{\bf{b}}}
\def\B{{\bf{B}}}
\def\a{{\bf{a}}}
\def\am{{{\bf{a}}_{\rm m}}}

\def\X{{\bf{X}}}
    \def\z{\bm{e}_z}
\def\tx{\bm{e}_{\tilde x}}  \def\ty{\bm{e}_{\tilde y}}
        
\def\S{{\bf{S}}}
\def\bOm{{\bm{\Om}}}
\def\half{{\textstyle \frac{1}{2} }}
\def\PhiBP{\Phi_{\rm BP}}

\newcommand{\bi}[1]{\mbox{\boldmath ${#1}$}}
\newcommand{\ignore}[1]{\relax}

\def\om{\Omega}
\def\Om{\omega}
\def\rmd{{\rm d }}
\def\rmi{{\rm i }}

\def\tP{t_{\rm P}}

\def\B{{\bf{B}}}

\def\a{{\bf{a}}}
\def\b{{\bf{b}}}
\def\J{{\bf{j}}}

\begin{document}

\title{Geometric nature of the environment-induced Berry phase
and geometric dephasing}

\author{Robert S. Whitney}
        \affiliation{
                Departement de Physique Th\'eorique, 
                Universit\'e de Gen\`eve, 
                24 quai Ernest-Ansermet, 
                1211 Gen\`eve 4, Switzerland.}
\author{Yuriy  Makhlin}
        \affiliation{Institut  f\"ur  Theoretische Festk\"orperphysik,
                Universit\"at   Karlsruhe,  D-76128   Karlsruhe, Germany}
        \affiliation{Landau Institute for Theoretical Physics, Kosygin st. 2,
                117940 Moscow, Russia.}
\author{Alexander Shnirman}
        \affiliation{Institut  f\"ur  Theoretische Festk\"orperphysik,
                Universit\"at   Karlsruhe,  D-76128   Karlsruhe, Germany.}
\author{Yuval Gefen}
        \affiliation{Department of  Condensed Matter  Physics, The Weizmann
                Institute of Science,  Rehovot  76100,  Israel.}
\date{January 26, 2005}


\begin{abstract}
We investigate the geometric phase or Berry phase (BP)
acquired by a spin-half which is both subject to a slowly
varying magnetic field and weakly-coupled to a dissipative environment
(either quantum or classical).  
We study how this phase is modified by the environment and 
find that the modification is of a geometric nature.
While the original BP (for an isolated system) 
is the flux of a monopole-field through the loop traversed by the magnetic 
field, the environment-induced modification of the BP
is the flux of a quadrupole-like field.
We find that the environment-induced phase is complex, and its    
imaginary part is a geometric contribution to dephasing. Its sign 
depends on the direction of the loop.
Unlike the BP, this geometric dephasing is 
gauge invariant for open paths of the magnetic field. 
\end{abstract}

\maketitle

{\it Introduction.}  The Berry phase (BP) is a fundamental
quantum-mechanical phenomenon related to the
adiabatic theorem.  Berry~\cite{Berry84} showed that the phase acquired by an
eigenstate of a slowly varying Hamiltonian $H(t)$ is related to the geometric
properties of the loop traversed by $H(t)$.  In the presence of dissipation the
condition of adiabaticity and the existence of the Berry phase require careful
analysis. The widespread criterion of adiabaticity, based on a comparison of 
the 
rate of change of the Hamiltonian with the gap in the
spectrum should be modified to involve the matrix elements of
the system-environment coupling, cf.~\cite{Whitney03-prl}.
Here we study the interplay of the varying field and the dissipation,
analyzing the BP in the limiting case of weak system-environment coupling.  
This
analysis is relevant to the recent and proposed experiments to 
manipulate quantum two-level systems (qubits).
Our findings could be tested in 
solid-state qubits, such as 
superconducting nanocircuits~\cite{Nakamura99,Vion02,Chiorescu03,Falci00}.

Berry~\cite{Berry84} considers a two-level spin-half system in a magnetic
field~\cite{footnote:units}, 
which is varied
slowly along a closed path:  $H_{\rm spin}= -\half \B(t) \bsigma$.
The rate of the field's change is characterized by the time to complete 
the loop, $t_{\rm P}$.  In the adiabatic limit, $B t_{\rm
P} \gg 1$, the relative phase acquired by the eigenstates is a sum $\Phi = \oint 
|\B(t)| dt + \PhiBP$ of the dynamical and Berry phases. The latter is 
geometric, it depends on the geometry of the loop but not on the
details of its traversal (for an isolated spin-half it is given by the 
solid angle subtended by the loop $\B(t)$). 
In the spin language, the evolution is a rotation of the spin
by an angle $\Phi$ about $\B$.

If the spin is not isolated, the dynamics are more complicated.  For a
static field, $\B$, dissipation induces energy and phase relaxation
processes (with the time-scales $T_1$, $T_2$ respectively) 
and a Lamb-like shift of the level splitting, $\delta B_{\rm Lamb}$, 
which modifies the dynamical phase.  
Strong dephasing masks the BP, however it can be observed if
$B \gg T_2^{-1}$~\cite{Whitney03-prl}.  In this weak-coupling
limit one can carry out a BP experiment slowly enough for non-adiabatic effects
to be ignored ($t_{\rm P} \gg B^{-1}$) while ensuring it is fast enough that
dephasing has not destroyed all phase information ($t_{\rm P} \lesssim T_2$).

\begin{figure}
\centerline{\hbox{\includegraphics[width=0.8\columnwidth]{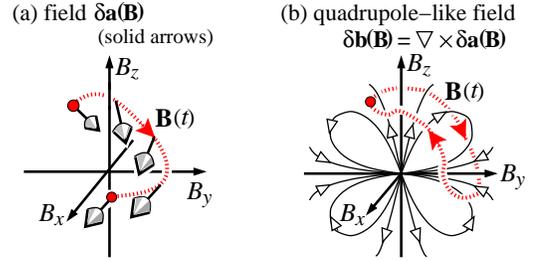}}}
\caption[]{\label{Fig:quadrupole}
(a) The field $\delta {\bf a}(B)$, see eq.~(\ref{Eq:delta_a}), 
in the vicinity of an open path of ${\bf B}(t)$ (the dotted line).
The environment-induced modification of the Berry phase, 
$\delta \Phi_{\rm BP}$, is the integral of $\delta {\bf a}(B)$ 
{\em along} this path.
For such open-paths, the geometric dephasing is gauge-invariant 
(unlike the phase).
(b) The quadrupole-like field $\delta {\bf b}(B)$, 
see eqs.~(\ref{Eq:delta_b_B}-\ref{Eq:delta_b_theta}),
($\delta {\bf b}(B)$ is cylindrically symmetric about the $B_z$-axis).
For a closed path of ${\bf B}(t)$, $\delta \Phi_{\rm BP}$
is given by the flux of $\delta {\bf b}(B)$ {\em through} the path.
}
\vskip -3mm
\end{figure}

In this letter we show that the coupling to the environment 
modifies the BP when the magnetic field is slowly varied.
In the adiabatic limit this modification, $\delta\PhiBP$,
is {\it geometric} (see Fig.~\ref{Fig:quadrupole}) and {\it complex}~\cite{Garrison1988}.
Its real part is an environment-induced BP, 
thus the total BP is $\PhiBP^{(0)}+ {\rm Re}[\delta\PhiBP]$
(where $\PhiBP^{(0)}$ is the BP for an isolated spin-half).
Its imaginary part, ${\rm Im}[\delta\PhiBP]$, 
is a geometric correction to dephasing, 
whose sign depends on the direction of the loop.
The magnitude of coherency will thus depend on the 
sign of the loop's winding number. Notably ${\rm Im}[\delta\PhiBP]$ 
is gauge-invariant for {\em open} as well as closed paths of the $B$-field.

Specifically we study the environment-induced BP 
for an arbitrary path $\B(t)$,
allowing us to analyze its geometric nature. For an isolated
spin-half the Berry phase is given by the flux through the closed loop $\B(t)$
of the field $\b_{\rm m}(\B)$ 
of a charge-one monopole at the origin in 
$\B$-space~\cite{Berry84}: we show that the environment-induced modification
may be interpreted
similarly and find the corresponding field distribution $\db(\B)$ in the
$\B$-space.  This field (and
thus $\delta\PhiBP$) scales quadratically with the spin-environment
coupling, similarly to the dissipative rates,  $T_1^{-1}$, $T_2^{-1}$.
If the environment-induced field fluctuates along a single direction, $z$, the
field $\db(\B)$ is axially symmetric, has no axial component, $\delta
b_\varphi=0$, and the angular distribution of a quadrupole.  
The field $\db(\B)$ is given by a sum over frequencies
of the environment modes (for the lowest non-trivial order in the strength
of the spin-environment coupling). 
The contribution of low frequencies $\om\ll B$ to $\db$ 
is exactly the field of a quadrupole, while high-frequency modes
have the angular dependence of a quadrupole 
but a different radial dependence.

Let us emphasize the novel points of our analysis in comparison to earlier work
on BP in systems coupled to quantum or classical environments 
\cite{Whitney03-prl,earlierBPdiss,GaitanAndMore,DeChiaraPalma}. 
Some of these works have focused on the visibility of BP in spite of 
the dephasing, 
they did not find the modification of the BP.
In \cite{earlierBPdiss} it is absent because
the Master equation used there  neglects the effect of
$\dot{\bf B}\neq 0$ on the dissipative rates.
A modification of the BP 
was found in  \cite{Whitney03-prl}, however the time-dependence of $\B(t)$ 
used there was too specific to see the geometric
nature of this modification.  Here we are able to consider a totally arbitrary 
(slowly varying) $B(t)$ and show, for the first time, that the modification to 
the BP is geometric. We also observe that
the effect of the BP on the dephasing rate is geometric.

To find the BP we analyze the (directly observable) phase factor in the 
evolution operator for a given field dynamics $\B(t)$.  
In this phase we attribute to BP the
contributions independent of $\tP$ \cite{footnote2}
(those $\propto \tP$ are part of the
dynamical phase, and the terms ${\cal O}(1/\tP)$ are non-adiabatic 
corrections~\cite{footnote3}).

For open paths the BP is gauge-dependent~\cite{WilczekZee},
we show that this is not the case for the geometric dephasing, 
${\rm Im}[\delta\PhiBP]$ . 
The gauge-dependence of the BP is a consequence of ambiguity in the
choice of instantaneous basis for a given ${\bf B}(t)$;
${\bf B}(t)$ defines the instantaneous $\tilde z$-axis but the $\tilde x$-axis may lie
anywhere in the plane perpendicular to ${\bf B}(t)$.
Gauge transformations rotate the $\tilde{x}$-axis in this plane. 
The ambiguity is absent for closed loops, where the
final basis must coincide with the initial one.
In addition one can monitor the magnitude of the transverse spin
component (dephasing); this magnitude 
is independent of the choice of $\tilde{x}$-axis,
i.e. the dephasing (and corrections to it) is gauge-invariant even for
open paths.

\begin{figure}
\centerline{\hbox{\includegraphics[width=0.3\columnwidth]{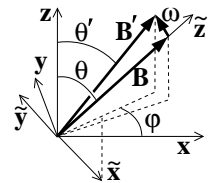}}}
\caption[]{\label{Fig:frames}
The laboratory $(x,y,z)$ and rotating $(\tilde x,\tilde y,\tilde z)$ frames.}
\vskip -3mm
\end{figure}

{\it Berry phase for an isolated spin.}
We evaluate the BP for an 
isolated spin-half~\cite{Berry84} using a rotating frame (RF)\cite{Berry87},
the one we choose (see Fig.~\ref{Fig:frames}) 
has its $\tilde z$-axis along the field $\B(t)$
and $\ty \perp\z$ (then $\tx(t_{\rm P}) = \tx(0)$).
In this frame the magnetic field is $\B'=\B+\bOm$, 
where the angular velocity of the RF is
\begin{equation}
\label{Eq:omega_inv}
\bOm = \dot\theta\, \ty + \dot \varphi\, \z
\ .
\end{equation}
Transforming to the RF gives the Hamiltonian, 
$\tilde{H} 
=  UH_{\rm spin}U^{-1} + i\dot U U^{-1}
=-\half (\tilde \B + \tilde \bOm)\hat{\bi{\sigma}}$,
where the transformation $U =
\exp(i\theta\hat\sigma_y/2)\, \exp(i\varphi\hat\sigma_z/2)$ 
(with $B$, $\theta$, $\varphi$ being spherical coordinates 
of $\B(t)$)\cite{footnote:Pauli-matrices};
$\tilde\B = (0,0,B)$ and 
$\tilde\bOm=(\Om_{\tilde x},\Om_{\tilde y},\Om_{\tilde z})=
(-\sin\theta\, \dot\varphi, \dot\theta, \cos\theta\, \dot\varphi)$ 
are coordinates in the RF.
The evolution in the RF is a rotation of the spin about the $\tilde z$-axis 
by the angle
$\oint dt |\B'(t)|$.  To lowest order in $\Om$ the accumulated phase is
$\int dt (B(t) + \Om_{\tilde z})$ 
\cite{footnote:RF}.  
Subtracting the dynamical phase,
we find the BP;
\begin{equation}
\label{Eq:Berry_Phase} \PhiBP =\int dt \,\Om_{\tilde z}(t)\ =
\oint d\varphi\,\cos\theta = \oint\! d\B \,\a(\B)\,.
\end{equation}
In the chosen frame (gauge) the `vector potential' $\a$ has only one non-zero 
spherical component $a_\varphi=1/(B  \tan\theta)$.
Stokes' theorem is used to writing eq~(\ref{Eq:Berry_Phase}) as  
\begin{equation}
\label{Eq:Berry_Phase_Monopol}
\PhiBP = 
\int {d\S}\,  \b(\B),\ \ \b(\B)\equiv {\bf\nabla}_\B\times \a(\B)
\,,
\end{equation}
where ${\bf\nabla}_{\B}$ denotes  derivatives w.r.t. the components of $\B$, and 
the field $\b(\B)$ is that of a
charge-one monopole, $b_B=1/B^2$.

{\it Environmental contribution to BP.}
Adding a noisy environment-induced field 
$\hat{\X}(t) \equiv \hat{X}(t) \z$, the Hamiltonian reads
(in the lab frame)\cite{footnote:units}
\begin{equation}
\label{Eq:Spin_X_Model}
\label{Eq:Ham}
\hat{H}= -\half {\bf B}(t) \hat{\bi{\sigma}}
-\half \hat{X} \hat{\sigma}_z + \hat{H}_{\rm env} (\hat{X}) \ .
\end{equation}
This Hamiltonian models a situation in which the 
spin-environment coupling is strongly {\em anisotropic}.
This is often true for solid-state qubits, where
various `spin' components couple to entirely different environmental
degrees of freedom~\cite{Nakamura99,Vion02,Chiorescu03}.
Our analysis is easily generalized for 
multi-directional coupling\cite{footnote4}.
Below we express our results in terms of the statistical
properties (correlators) of the environment's noise, $\hat{X}(t)$.
We consider any environment which gives rise to 
Markovian evolution, 
i.e. $\langle \hat{X}(t)\hat{X}(0)\rangle$ decays
on a timescale $\tau \ll t_{\rm P},T_{\rm diss} \equiv 
{\rm min}(T_1,T_2,\delta B_{\rm Lamb}^{-1})$ however we do {\em not} assume $\tau \ll B^{-1}$. 
We further consider weak dissipation such that
$BT_{\rm diss}\gg 1$.
In the RF the Hamiltonian reads:
\begin{equation}
\label{Eq:Diss_Rotating_Frame} 
\tilde H       =
-\half B(t)  \hat{\sigma}_{z} -
\,\half \tilde\bOm(t)\hat{\bsigma}  -  \half \hat{X} (\cos\theta
\hat{\sigma}_{z} -\sin\theta \hat{\sigma}_{x})  
+\hat{H}_{\rm env}\,.
\end{equation}
We analyze this system in the eigen-basis of 
$B(t)  \hat{\sigma}_{z} + \tilde\bOm(t)\hat{\bsigma}$, 
with the total field $B'(t) \approx B(t) + \omega_{\tilde z}$
and the new angle $\theta'$ as defined in Fig.~\ref{Fig:frames}.

The small terms $\propto\Om, \langle \hat{X}^2\rangle$ 
induce corrections to the evolution.  As
discussed above, the correction $\propto\Om$ to the rate of phase
accumulation yields the BP, while the correction 
$\propto \langle \hat{X}^2\rangle $ gives $T_1^{-1},T_2^{-1},
\delta B_{\rm Lamb}$.  
Here we evaluate 
the environment-induced BP $\propto \langle \hat{X}^2\rangle\Om$. 
To this end, we reduced the problem
with a time-dependent field to a problem with a stationary field
(to leading order \cite{footnote:RF}) by
going to the RF and then use the standard formalism for evaluation of
dissipative effects. 

Consider the kinetic equation for the reduced density  matrix of the 
spin $\rho$.
Iteration of the Liouville 
equation~\cite{Bloch_Derivation,Redfield_Derivation,textbooks,
Schoeller_PRB,MSS-review}
leads to a Dyson-type master equation:
\begin{equation}
\label{Eq:Dyson_Master_Equation}
\dot\rho_{ij}(t)= \frac{\rmi}{2}\left[B'\sigma_z, \rho(t)\right]_{ij}
\!+\! \int_0^t \!\!\! dt_1 \Sigma_{ij,i'j'}(t,t_1) \rho_{i'j'}(t_1) \ .
\end{equation}
The `self-energy' $\Sigma (t,t_1)$ can be evaluated perturbatively in $\hat{X}$;
diagrammatically it is the sum of irreducible diagrams
\cite{Schoeller_PRB,MSS-review}.
For short-correlated noise $\tau \ll T_{\rm diss}$, 
one can make a Bloch-Redfield approximation 
\cite{Bloch_Derivation,Redfield_Derivation}
giving Markovian evolution.
We need only consider the off-diagonal density matrix element, 
$\rho_{12}(t)$, as it contains all phases (and dephasing) .  
After a secular approximation 
(for $\Gamma \ll B'$)~\cite{chapter_Landau_Lif}, 
it is given by \cite{footnote-limit-on-integral}
\begin{eqnarray}
&\dot\rho_{12}(t)=\left[\rmi B'(t)+ \Gamma(t)\right]\,\rho_{12}(t)
\,,&
\label{Eq:Bloch_Redfield}
\\
&\Gamma(t) \equiv \int\nolimits_{-\infty}^{t}
dt_1\; \Sigma_{12,12}(t,t_1)\; \exp\left(-\rmi\int\nolimits_{t_1}^{t} 
B'(\tau)d\tau\right)\,.&
\end{eqnarray}
We follow~\cite{Bloch_Derivation,Redfield_Derivation,Schoeller_PRB,MSS-review}
evaluating $\Sigma_{12,12}$ 
to lowest order ($2^{\rm nd}$-order) in $\hat{X}$.
This ``golden-rule'' calculation yields 
\begin{eqnarray}
\label{Eq:Gamma}
\Gamma(t)&=&-\int\nolimits_{-\infty}^{t} dt_1 S(t-t_1)
\Big[  \cos\theta'(t)\cos\theta'(t_1)
\nonumber \\
& & +{\textstyle \frac{1}{2}} \sin\theta'(t)\sin\theta'(t_1) 
\exp \left({-\rmi{\textstyle \int_{t_1}^{t}} B'(\tau)d\tau}\right) 
\Big] \ , \qquad
\end{eqnarray} 
where 
$S(t-t_1) \equiv 
(1/2)\left\{\langle \hat{X}(t)\hat{X}(t_1)\rangle  + 
\langle \hat{X}(t_1)\hat{X}(t)  \rangle\right\}$ is the symmetrized  
environment correlator~\cite{StPeter_Berry}.
If we neglect all order-$\omega$ effects, we obtain
\begin{eqnarray}
\label{Eq:Gamma_0}
\Gamma_{0}(t)=
-\frac{\rmi}{2}
\int \! \frac{d\om}{2\pi}\,S(\om) \left[
\frac{\sin^2\theta(t)}{\om-B+\rmi 0^+} + \frac{2\cos^2\theta(t)}{\om+\rmi 0^+}
\right]\quad
\end{eqnarray}
where the integral is along the real axis and the function 
$S(\om)$ is the Fourier transform of $S(\tau)$, it is real and even.
The real part of $\Gamma_{0}$ is the dephasing rate,
which is given by the residue of the poles at $\om=0,B$.
The imaginary part of $\Gamma_{0}$ is the Lamb shift,
coming from the principal value of the integral 
(only the $\sin^2\theta$ term contributes).

Now we take into account the order-$\omega$ term in $\B'(t)$
and the time-dependence of $\B'(t)$; variations
of the angle between $\B$ and $\X$ and of 
$B=|\B|$ [i.e. $\theta(t_1)\ne\theta(t)$ and $B(\tau)\ne B(t)$ in
(\ref{Eq:Gamma})].
To lowest order in $\Om$ the modification of the 
rate (\ref{Eq:Gamma_0}) are:
\begin{eqnarray}
\label{Eq:delta_Gamma_1}
\delta_\varphi \Gamma(t)
&=& 
\rmi \dot\varphi \,   \sin^2\theta(t)\cos\theta(t)
\times F(B)
\,,
\\
\nonumber
\delta_\theta \Gamma(t)&=&
\rmi \dot\theta \,  \sin\theta(t)\cos\theta(t) \times G(B)
\,,
\\
\nonumber
\delta_B \Gamma(t)&=& 
\rmi \dot B\, \sin^2\theta(t) \times {\textstyle \frac{1}{2}} G'(B)
\,,
\end{eqnarray}
the dot indicates $(\rmd / \rmd t)$, 
the prime indicates $(\rmd / \rmd B)$, and
\begin{eqnarray}
\label{Eq:F(B)}
F(B) 
&\equiv& 
\rmi\frac{S(0)}{B} - \frac{1}{2}
\int\frac{d\om}{2\pi}\,\frac{S(\om)
\left(3B-2\om\right)}{B(\om-B+\rmi 0^+)^2}
\,,
\\
G(B) 
&\equiv& 
\frac{\rmi}{2} \int\frac{d\om}{2\pi}
\left[ \frac{S(\om)}{(\om-B+i0^+)^2}-\frac{2S(\om)}{(\om+\rmi 0^+)^2} \right] 
\,.
\nonumber 
\end{eqnarray}
From Eq.~(\ref{Eq:Bloch_Redfield}) one sees that these ${\cal O}[\Om]$-terms 
in $\Gamma(t)$ generate ${\cal O}[\Om^0]$-terms in the total phase acquired by
$\rho_{12}$:
$\delta \PhiBP =  \int [
F(B)\sin^2\theta\cos\theta\,
d\varphi + G(B)  \sin\theta\cos\theta\,
d\theta $ $+ \frac{1}{2} G'(B)\sin^2\theta\, dB ]$.
This is geometric, and corresponds to
$\delta \PhiBP =  \int \rmd \B \, \delta \a(\B)$
(cf. eq.~(\ref{Eq:Berry_Phase})) with the complex `vector potential' 
\begin{eqnarray}
\label{Eq:delta_a}
&\delta a_{\varphi} = B^{-1}F(B) \sin\theta \cos\theta \,,&\\
&\delta a_{\theta}  = B^{-1}G(B) \sin\theta \cos\theta \,,
\quad \delta a_{B} = {\textstyle \frac{1}{2}} G'(B) \sin^2\theta \,,&
\nonumber 
\end{eqnarray}
For open-paths of ${\bf B}(t)$ this is our main result,
the imaginary  part of this field (which gives geometric dephasing) 
is gauge-{\it independent}, while the real part is not.
For closed paths of ${\bf B}(t)$ we use Stokes' theorem to 
write  $\delta \Phi_{\rm BP}=\int {d\S}\,  \delta\b(\B)$
(cf.~eq.~(\ref{Eq:Berry_Phase_Monopol})),
finding that 
$\delta \b(\B)\equiv {\bf\nabla}_{\B} \times {\delta \a}$  
has two non-zero components;
\begin{eqnarray}
\label{Eq:delta_b_B}
\delta b_{B} &=& B^{-2} F(B)(3\cos^2\theta - 1)
\,,\label{bB}
\\
\label{Eq:delta_b_theta}
\delta  b_{\theta} &=& - B^{-1} F'(B) \sin\theta\cos\theta
\,.
\label{btheta}
\end{eqnarray}
Thus it is independent of 
$\delta a_{\theta}$ and $\delta  a_{B}$, 
their only role (for closed paths of $\B(t)$) is to form a `pure gauge'; 
this is related to a symmetry discussed below.
The angular dependence of $\db$ resembles that of a  quadrupole.
For slow environment modes ($\om \ll B$), ${\rm Re} [F(B)] \propto B^{-2}$ and 
hence  $\db(\B)$ is a quadrupole field.  
For other environment modes, $\db(\B)$ has non-zero curl and
zero divergence.  Thus it is not a sum multipoles,
it is the field generated by a  pseudo-current, 
$\J(\B) \equiv{\bf\nabla}_{\B}\times \b(\B)$, 
with one non-zero component
\begin{eqnarray}
\label{eq:J_phi}
j_\varphi (B) = B^{-3}\big(6F(B) - B^2F''(B) \big) \sin
\theta \cos \theta\,.
\end{eqnarray}
\ignore{
For a slow environment $j_\varphi$ is singular at $B=0$, i.e. a quadrupole.
Note that if $|\B|$ is kept constant, one does not notice the 
non-quadrupole $\delta b_\B$, and $\db$ appears to a quadrupole.
}
Since we can ignore $\delta a_\theta$ and $\delta a_B$ for closed paths,
eq.~(\ref{Eq:delta_a}) leads to the following pretty result:
the total (complex) BP for a closed path is
\begin{equation}
\PhiBP 
= \oint \rmd \varphi \ \frac{\rmd}{\rmd {B_z}} (B +\delta B_{\rm Lamb} 
+ \rmi T_2^{-1}) \ .
\label{Eq:Phi=d/dBz}
\end{equation}
This result can be understood as follows. 
For a time-independent $\B$ the acquired phase is 
$\int dt\, {\cal E}$, where ${\cal E}$
is the term in parentheses in Eq.~(\ref{Eq:Phi=d/dBz}).
When  $\B$ is time-dependent, 
$\B \rightarrow \B + \bOm$ in the RF
changing the phase by $\int dt\, \bOm {\bf\nabla_B} {\cal E}$;
with Eq.~(\ref{Eq:omega_inv}) this gives Eq.~(\ref{Eq:Phi=d/dBz}).

{\it Symmetry considerations.}
We now show that the vanishing of $b_{\varphi}$ follows from 
the symmetries of the problem.
Firstly, under time-reversal $\B(t),\X(t)\to -\B(-t),-\X(-t)$
the BP between the excited and ground states is
invariant, while in eq.~(\ref{Eq:Berry_Phase_Monopol}) $d\S\to-d\S$. This
implies that ${\bf b}(-\B, -X) = -\b(\B, X)$.
Secondly, if we instead rotate all spins and
fields by angle $\pi$ about the axis
$\hat\varphi$ perpendicular to both $\B$ and $X$ (e.g. $\B\to-\B$), the field
$\b$ would also rotate. Thus, $\b(-\B,-X)=R^\pi_{\varphi} \b(\B,X)$.
To satisfy both equalities, $b_\varphi \equiv 0$.

For isotropic coupling, the Hamiltonian's rotational symmetry
guarantees that $b_\varphi=b_\theta=0$. 
The monopole (cf.~eq.~(\ref{Eq:Berry_Phase_Monopol})) is quantized so
$\delta b_B=0$; hence $\delta \Phi_{\rm BP}=0$
to all orders in the coupling to the environment.

{\it Contribution of the slow modes} ($\om \ll B$).
Here $F(B)\propto B^{-2}$, so $\delta {\bf b}$ is a quadrupole field. 
This effect of (for instance, classical) slow modes
can be understood
by noting that $(\B(t)+\X(t))$ varies adiabatically, hence 
\begin{equation}
\PhiBP = \oint d(\B+\X) \am(\B+\X)
=\oint d\B \am(\B+\X)
\,,
\label{Eq:expansion}
\end{equation}
where $\am$ is the field of a charge-one monopole at the origin.
We disregard boundary corrections due to $\X(\tP)\ne\X(0)$ and 
drop the term $\oint d\X\, \am(\B+\X)$ \cite{footnote5}.
The remaining term is the field of a monopole at the point $-\X$. In the 
multipole expansion,
$\am(\B+\X)=\am(\B) +
X_\alpha \nabla_\alpha \am(\B)
+
\half X_\alpha X_\beta
\nabla_\alpha \nabla_\beta \am(\B) + \cdots$.
The first term produces the unperturbed BP,
the second (dipole) term vanishes after averaging over the
fluctuations \cite{DeChiaraPalma}.
Thus the
quadrupole term gives the leading environment-induced modification of the BP.
For noise along the $z$-axis this term reads
$D_{\alpha\beta}
\nabla_\alpha\nabla_\beta \am/6$, with the
quadrupole moment $D_{\alpha\beta} =
\langle X^2\rangle \mathop{\rm diag}(-1,-1,2)$.

{\it Concerning experiments.} Our results imply that the traversal of the 
same loop
$\B(t)$ in opposite directions would yield, apart from different BP's,
different dephasing. We further note that a noise spectrum suppressed
at $\om=0$, $B$ will increase $T_2$ while having little 
effect on $\delta \Phi_{\rm BP}$, thus aiding the 
observation of $\delta \Phi_{\rm BP}$.

This work is part of CFN (DFG), supported in part by
the Minerva (DFG) Foundation, the EC-RTN Spintronics, 
and ISF of IAS.
RW, YM and YG were supported by the EPSRC, the Dynasty Foundation and 
the AvH Foundation (Max-Planck award) respectively. 
The Transnational Access prog.~supported the 
\hbox{WIS visit of YM,AS.}

\bibliographystyle{apsrev}

\begin{thebibliography}{17}
\bibitem{footnote:units}
        All energies and magnetic fields are in units of inverse time
        ($g\mu_{\rm B}=\hbar=1$).
\bibitem{footnote2}
Here we ignore the boundary terms, $\Phi_{\rm shift}$; 
cf.~\cite{Whitney03-prl}.
\bibitem{footnote3}
        If $\B(t)$ traverses a loop $n$ times in a time $t_{\rm P}\sim T_2$, \
        the leading non-adiabatic correction  
        $\sim n^2 \langle \hat{X}^2 \rangle$,
        while $\delta \Phi_{\rm BP}\sim n \langle \hat{X}^2 \rangle$.
        We can separate the two by varying $t_{\rm P}$
        (keeping $t_{\rm P}\sim T_2$) or $n$.
        Alternatively the
        environment's spectrum may make non-adiabatic effects 
        $\ll {\rm Re}\left[\delta\PhiBP\right]$, 
        because $\delta\PhiBP$ comes from the whole spectrum while $T_2^{-1}$
        is dominated by frequencies $0,B$ (see below eq.~(\ref{Eq:Gamma_0})).
\bibitem{footnote:Pauli-matrices}
  To avoid confusion we define $\hat{\bi{\sigma}}\equiv 
  (\hat\sigma_x,\hat\sigma_y,\hat\sigma_z)$ as the same Pauli
  matrices, e.~g.~$\hat\sigma_z = {\rm diag}(1,-1)$, in all frames.
\bibitem{footnote:RF}
Typically $(\tilde{\B}+\tilde\bOm)$ in the RF is non-stationary and is not 
quite parallel to the $\tilde z$-axis. This only gives non-adiabatic 
corrections, ${\cal O}(\omega^2/B,\dot{\omega}/B)$;
cf. \cite{Berry87}.
\bibitem{footnote4}
For multi-directional coupling $\delta \Phi_{\rm BP}$ is the sum of the
$\delta \Phi_{\rm BP}$s calculated separately for each coupling axis 
to ${\cal O}[\langle \hat{X}^2 \rangle]$.  
We find  $\delta \Phi_{\rm BP}=0$ for isotropic coupling.
\bibitem{footnote-limit-on-integral}
The above integral is dominated by $t_1$ close to $t$ 
thus we can take the lower bound of integration to $-\infty$ 
with little effect 
(except at small $t$ where it causes a small ``preparation'' correction).
\bibitem{footnote5}
Since for $\X\parallel\z$ in the chosen gauge $d\X \perp \am$; 
this also holds for general $\X$ as long as $\X(t)$ and $\X(-t)$ have equal
weights since their contributions cancel.
\end{thebibliography}

\end{document}